\documentclass[trackchanges,twocolumn]{aastex701}
\usepackage{graphicx}
\usepackage{braket}
\usepackage{url}
\usepackage[toc,page]{appendix}
\usepackage{physics}
\usepackage[thinc]{esdiff}
\usepackage{subcaption} % Required for subfigure
\usepackage{comment}
\usepackage{bm}

\usepackage[dvipsnames]{xcolor}
\usepackage{tensor}
\usepackage{esdiff}
\usepackage{amsmath, amssymb}
\usepackage{mathrsfs}  
\usepackage[linewidth=1pt]{mdframed}

\usepackage{hyperref}

\renewcommand{\d}{{\rm d}}
\newcommand{\e}{\text{e}}
\newcommand{\uvec}[1]{{\bm{\hat{{#1}}}}}
\newcommand{\un}{{\bm{\hat{{n}}}}}

\newcommand{\rot}{{\rm R}}

\DeclareMathOperator{\bi}{\mathbb{{Y}}}
\DeclareMathOperator{\tri}{\mathbb{{T}}}
\DeclareMathOperator{\cg}{\mathcal{{C}}}

\newcommand{\sfigref}[2]{Fig.~\hyperref[#1]{\ref{#1}#2}}

\renewcommand{\abs}[1]{\left| #1 \right|}

\begin{document}
\nolinenumbers
\title{Capturing statistical isotropy violation with rotational averages}

\author[orcid=0009-0004-7856-9587,sname='R']{Vaishali R}
\affiliation{Raman Research Institute, C.~V.~Raman Avenue, Sadashivanagar, Bengaluru 560080, India}
\email{vaishali@rrimail.rri.res.in}  

% \correspondingauthor
\author[orcid=0000-0002-5267-1944,sname='Mukherjee']{Dipayan Mukherjee} 
\affiliation{Raman Research Institute, C.~V.~Raman Avenue, Sadashivanagar, Bengaluru 560080, India}
\email[show]{dipayan@rrimail.rri.res.in}

\author[orcid=0009-0002-2757-1957,sname='Tyagi']{Akshank Tyagi}
\affiliation{Raman Research Institute, C.~V.~Raman Avenue, Sadashivanagar, Bengaluru 560080, India}
\email{akshank101@gmail.com}

\author[orcid=0000-0003-3764-8102,sname=Souradeep]{Tarun Souradeep}
\altaffiliation{Inter University Centre for Astronomy and Astrophysics, Post Bag 4, Ganeshkhind, Pune-411007, India}
\affiliation{Raman Research Institute, C.~V.~Raman Avenue, Sadashivanagar, Bengaluru 560080, India}
\email{tarun@rri.res.in}

\correspondingauthor{Dipayan Mukherjee}

\begin{abstract}
Recent high precision cosmological observations have revealed several anomalies in the Cosmic Microwave Background (CMB), indicating possible violations of statistical isotropy (SI) --- rotational invariance of statistical measures of a random skymap. The Bipolar Spherical Harmonic (BipoSH) formalism, a widely used complete representation of both the SI and non-Statistically Isotropic (nSI) components of a random skymap, is defined and implemented in the spherical harmonic space.
In this work we probe the rotational symmetry violation of the correlation function using weighted rotational averages, providing a geometric interpretation of BipoSH as real space operations. We establish that the BipoSH measures are related to averages of the correlation function over all rotated configurations weighted by Wigner matrices.
% that systematically captures the breakdown of SI at different scales.  
As a demonstration, we consider a phenomenological dipole modulation model and numerically implement the rotational average measures and show their agreement with their harmonic space counterparts. The real space approach could be advantageous in formulating BipoSH like nSI measures on partial sky. We envisage broader applications of this framework to the studies of primordial non-Gaussianity, CMB polarisation, weak gravitational lensing and the characterisation of general random fields on a sphere.
\end{abstract}




\section{Introduction} 

The cosmological principle postulates that the universe is homogeneous and isotropic on sufficiently large scales. Observations of the cosmic microwave background (CMB) and the large-scale structure support this ansatz, indicating a uniform universe across hundreds of megaparsecs~(\cite{Wu:1998ad}). Extended to cosmological perturbations, the principle requires the underlying stochastic field to be statistically homogeneous and isotropic, so that its correlation functions are invariant under spatial translations and rotations. For Gaussian fluctuations, this assumption collapses the entire statistical content of the CMB temperature anisotropies onto the angular power spectrum~(\cite{planck18_VII}).

As CMB observations have become increasingly precise, several features in the CMB sky, identified over the past two decades, suggest possible deviations from statistical isotropy (SI), collectively referred to as \emph{CMB anomalies}~(\cite{Schwarz_2016, planck18_VII, Planck:2018nkj}). They could indicate a fundamental violation of homogeneity and isotropy in the physics of the early universe~(\cite{Prunet:2004zy, Gordon_2007, Carroll:2008br, Bartolo_2012, Rath_2015, Kothari_2016, Ragavendra_2025, Aluri:2022hzs}), or they may be artefacts of instrumental and observational systematics~(\cite{Mitra:2004nx, Joshi:2009mj, Das:2014awa, Mukherjee:2013kga, Kumar:2014nda, Dipanshu:2023tix}). It is therefore imperative to systematically quantify statistical anisotropy (nSI) in CMB, both to discern its origin and to identify possible signatures of non-trivial physics in the very early universe.

Quantifying general nSI in CMB is non-trivial for several key reasons. The CMB correlation functions are defined on the sphere and thus require a harmonic decomposition adapted to the spherical geometry --- less straightforward than its three-dimensional Euclidean counterpart. Further, for a Gaussian field, SI imposes a unique structure on the correlation functions while its violation does not: isotropy can be broken in infinitely many ways~(\cite{Hajian:2003qq, Varshalovich:1988ifq}). Consequently, extracting a physical notion of the `degree of anisotropy' remains challenging. Finally, CMB observations provide but a single realization of the underlying random process, a fundamental obstacle to statistical inference.

The standard route to a fully general treatment of CMB nSI is the Bipolar Spherical Harmonic (BipoSH) basis~(\cite{Hajian:2003qq, Hajian:2005jh, Varshalovich:1988ifq}), which recasts anisotropy in the two-point function using the language of angular-momentum addition. The multipole structures of the two line-of-sight directions are coupled into a state of total angular momentum $L$, with $L = 0$ recovering the usual power spectrum and $L > 0$ encoding nSI. %Although mathematically robust, this harmonic construction comes at a price: it tends to obscure the geometric and physical meaning of the measures on the sky. Further, it can be observationally costly --- capturing even a large-scale, small-$L$ feature generally requires BipoSH coefficients up to high internal multipoles~\cite{Dipanshu:2023tix}.

In this paper we introduce a complementary real-space framework. Because the $N$-point correlation functions are rotationally invariant under the hypothesis of SI, the natural real-space diagnostic is the average of the correlation function over all rotated configurations. The key ingredient is the choice of weight: by weighting this rotational average with Wigner matrices, we systematically isolate the nSI features at each Bipolar $L$. We then map these rotational averages onto the BipoSH coefficients. This serves two ends. First, it endows the BipoSH formalism with a transparent, real-space geometric interpretation of how it encodes nSI. Second, it provides a new framework of extracting nSI signatures directly from real-space temperature data. We further extend the construction to the angular three-point correlation function in Appendix~\ref{appendix}.

The real-space approach carries advantages: it isolates the nSI signal at a given multipole without the full machinery of harmonic-space decomposition. Moreover, operating natively in real space sidesteps the partial-sky artefacts --- mode coupling and leakage --- that inevitably afflict harmonic analysis of the masked CMB sky. 

\section{BipoSH expansion}
A general two-point function $C(\un_1,\un_2)$ is suitably expanded in the BipoSH basis, which couples the two internal ranks ($\ell_1, \ell_2$) corresponding to the two directions ($\un_1, \un_2$) into a state of total angular momentum $L$~(\cite{Varshalovich:1988ifq}),
\begin{align}
\label{eq:2pcf_biposh}
C(\un_1, \un_2) 
= 
\sum_{LM} \sum_{\ell_1 \ell_2}
A_{\ell_1 \ell_2}^{LM} \bi_{\ell_1 \ell_2}^{LM}(\un_1, \un_2),
\end{align}
where $A_{\ell_1 \ell_2}^{LM}$ are the BipoSH coefficients; the BipoSH basis functions $\bi_{\ell_1 \ell_2}^{LM}(\un_1, \un_2)$ are defined in terms of tensor product of two spherical harmonics
\begin{align}
\begin{split}
\bi_{\ell_1 \ell_2}^{LM}(\un_1, \un_2) &\equiv \{Y_{\ell_1}(\un_1) \otimes Y_{\ell_2}(\un_2)\}_{LM}\\ &= \sum_{m_1 m_2} \mathcal{C}_{\ell_1 m_1 \ell_2 m_2}^{LM} Y_{\ell_1 m_1}(\un_1) Y_{\ell_2 m_2}(\un_2),
\end{split}
\end{align}
where $\cg^{LM}_{\ell_1 m_1 \ell_2 m_2}$ are the Clebsch-Gordan coefficients. The triangle inequality $|\ell_1 - \ell_2| \le L \le \ell_1 + \ell_2$ and the condition $M = m_1 + m_2$ are assumed in this expansion.

The coefficients $A_{\ell_1 \ell_2}^{LM}$ fully encode the statistics of the two-point function in the harmonic space. For a Gaussian field they therefore encode the full statistical content of the field, whereas for a non-Gaussian field the higher-point functions carry independent
information.

However, the nSI information is often spectrally spread out in the harmonic and bipolar-harmonic multipole space. Typically, one looks for nSI features in the correlation function at a given scale, characterized by the total multipole $L>0$.
Consequently, capturing the nSI signal for even a small $L$ mode (large scale) requires evaluating these coefficients up to arbitrarily high internal ranks. Most importantly, although mathematically robust, the abstract concept of angular momentum coupling obscures the geometric nature of the field wherein the underlying symmetry is broken in the case of nSI. We address this by presenting a real-space framework, specifically in the context of CMB temperature anisotropies.
% {\color{red} which can be extended to spin fields on the sphere.}

\section{Rotational averages of correlation functions}
Let $ C(\un_1, \un_2, \ldots, \un_N) $ be an arbitrary $N$-point function on $S^2 \times \cdots \times S^2$, where $\un_i$ are unit vectors on a sphere. The function is isotropic if it is invariant under arbitrary rotations of its arguments,
\begin{align}
C(\un_1, \un_2, \ldots, \un_N) = C(\rot \un_1, \rot \un_2, \ldots, \rot \un_N),
\end{align}
for all $\rot \in SO(3)$.
 We define the average of $C$ over all rotated configurations as
\begin{align}
\begin{split}
I(\un_1, \un_2, \ldots, \un_N)
=& \int_{SO(3)} \d \mu(\rot)\, W(\rot)\,\\
&\times C(\rot \un_1, \rot \un_2, \ldots, \rot \un_N),
\end{split}
\end{align}
where $W(\rot)$ is a chosen rotation-dependent weight, and $\d \mu(\rot)$ is the normalized Haar measure for the integration over the the rotation group $SO(3)$~(\cite{Varshalovich:1988ifq}). If $C$ is isotropic, $I$ becomes proportional to $C$. For an nSI $C$, its rotated configurations differ, and the weighted average probes the sensitivity of $C$ to rotations thus providing a measure of the `degree of anisotropy' in $C$, where the choice of $W(\rot)$ determines how this information is extracted. 

To arrive at suitable weight functions, it is instructive to consider the harmonic representations of these correlation functions. A generic one-point function $f(\un)$ is expanded in the standard spherical harmonic basis,
$
f(\un) = \sum_{\ell m} f_{\ell m} Y_{\ell m}(\un),
$
while the harmonic expansion of a general two-point function is given in the BipoSH basis in Eq.~\eqref{eq:2pcf_biposh}.
Importantly, a spatial rotation $\rot \in SO(3)$ transforms these harmonic bases linearly via the Wigner-$D$ matrices~(\cite{Varshalovich:1988ifq}): i.e., the spherical harmonics of multipole $\ell$ transforms as
\begin{align}
Y_{\ell m}(\rot \un) = \sum_{m'} D^\ell_{m'm}(\rot) Y_{\ell m'}(\un),
\end{align}
whereas the BipoSH basis state of total angular momentum $L$ transforms analogously
\begin{align}
\label{eq:biposh_rot}
\bi_{\ell_1 \ell_2}^{LM}(\rot \un_1, \rot \un_2) = \sum_{M'} D^L_{M'M}(\rot) \bi_{\ell_1 \ell_2}^{LM'}(\un_1, \un_2).
\end{align}
It is important to note that the Wigner-D matrix mixes the azimuthal indices $M$, while preserving the angular structure indicated by $L$. Thus if a two-point function contains an $L$-th nSI correlation, its behavior under rotations is entirely dictated by the Wigner matrix. 
Therefore, a non-zero integral with a weight constructed from $D_{MM'}^{L}(R)$ would indicate the presence of symmetry violation in the correlation function at the $L$-th scale.
 Note that these matrices are orthogonal with respect to the normalized integration measure $\d \mu(\rot)$ over $SO(3)$~(\cite{Varshalovich:1988ifq}),
\begin{align}
  \label{eq:wigner_ortho}
\begin{split}
\int \d \mu(\rot) D^{L_1}_{M_1 M_1'}(\rot) D^{L_2 \ast}_{M_2 M_2'}(\rot)
=&  \frac{1}{2L_1 + 1} \\
&\times \delta_{L_1 L_2} \delta_{M_1 M_2} \delta_{M_1' M_2'}.
\end{split}
\end{align}
With this motivation, we now construct the weighted rotational averages for the CMB two-point correlation function.

\subsection{Rotational averages of CMB two-point correlation function}
In the following, we denote $C(\un_1, \un_2) = \Braket{\Delta T(\un_1) \Delta T(\un_2)}$ to be the CMB correlation function, although the arguments shall remain valid for arbitrary two-point functions. 
In general, $C(\un_1, \un_2)$ is a real function defined on $S^2 \times S^2$.
We now consider different weight functions based on Wigner matrices to extract nSI information of the correlation functions at varying levels of detail.
\subsubsection{Uniform weight}
If we choose the weight $W$ to be isotropic ($\propto D^0_{00}$), i.e., independent of rotations, then the correlation function is uniformly averaged over all rotations. 
In this case, $I(\un_1, \un_2)$ captures the \emph{statistically isotropic (SI) part} of the correlation function through real space (\cite{Szapudi:2001qj}). That is, setting $W(\rot) = 1$,
\begin{align}
    I(\un_1, \un_2) 
    =& I_{\rm uni}(\un_1 \cdot \un_2)\\
    =& \int \d\mu(\rot) C(\rot\un_1, \rot \un_2)\\
    =& \sum_{\ell} (-1)^{\ell}
    \frac{\sqrt{2\ell+1}}{4\pi}
    A^{00}_{\ell\ell}
    P_\ell(\cos \theta).
\end{align}
$I_{\rm uni}(\un_1 \cdot \un_2)$ is a function of the relative angle $\theta = \cos^{-1}(\un_1 \cdot \un_2)$ and $A^{00}_{\ell \ell}$ are the BipoSH coefficients.
The coefficient in the Legendre expansion can be identified with the angular power spectrum $C_\ell$ as
\begin{align}
    A^{00}_{\ell \ell} = (-1)^\ell \sqrt{2 \ell + 1} C_\ell.
\end{align}
It is evident from above that the uniform rotational average is simply the isotropic angular
correlation function,
\begin{align}    
\label{eq:I_uni}
I_{\rm uni}(\un_1 \cdot \un_2)
= \sum_{\ell} \frac{2\ell+1}{4\pi} C_{\ell} P_{\ell}(\cos\theta)
= C(\cos\theta).
\end{align}

\subsubsection{Character function as weight}
To extract information of nSI in $C(\un_1, \un_2)$ \emph{at the multipole level}, we use the character function of the rotation group~(\cite{Varshalovich:1988ifq})
\begin{align}
\chi^L(\rot) 
= \sum_{M=-L}^{L} D^L_{MM}(\rot)
\end{align}
 as the weight in the rotational average, i.e.,
\begin{align}  \label{I_L}
    I^{L}(\un_1,\un_2)
    = 
    \int \d  \mu(\rot) \chi^L(\rot)
    C (\rot \un_1, \rot \un_2).
\end{align}
The quantity $I^{L}(\un_1,\un_2)$ captures the violation of isotropy by the two-point correlation function at the $L$-th multipole corresponding to the line-of-sight directions $\un_1$ and $\un_2$.
We can further average the squared absolute quantity $\abs{I^{L}(\un_1,\un_2)}^2$ over both the directions to arrive at a positive definite global measure of anisotropy
\begin{align}
    \kappa^L
    \equiv \int \frac{\d \Omega_{\un_2}}{4 \pi}
    \int \frac{\d \Omega_{\un_1}}{4 \pi}
    \abs{I^{L}(\un_1,\un_2)}^2.
\end{align}
% Note that the direction averages of un-squared $I^{L}(\un_1,\un_2)$ shall be $\propto \delta_{L0}$, thus can  only capture the isotropic contribution.
Note that it is essential to average the \emph{squared} quantity over the directions. The
direction average of $I^{L}(\un_1, \un_2)$ itself is proportional to $\delta_{L0}$, by the
orthogonality of the BipoSH basis, and would therefore retain only the isotropic contribution.

These rotational and spatial averages are fundamentally related to the BipoSH formalism as we will see. The direction-averaged measure $\kappa^L$ takes the form in the harmonic space
\begin{align}
   \kappa^L 
   = \frac{1}{(4 \pi)^2(2L+1)^{2}}
     \sum_{M=-L}^{L}
     \sum_{\ell_1 \ell_2}
    \abs{A^{LM}_{\ell_{1}\ell_{2}}}^{2}.
\end{align}
The quantity $\kappa^L$ can be identified as the Bipolar Power Spectrum (BiPS) introduced in~(\cite{Hajian:2003qq, Hajian:2005jh}). However, the present definition incorporates normalization for both the directional and rotational averages, which accounts for the extra prefactor appearing in our expression.

As the character function is the trace of the Wigner matrix, the azimuthal index $M$ is traced over in the above expressions. Consequently, the choice of this weight function averages out the spatial orientation of the multipole.

\subsubsection{Diagonal elements of Wigner matrices as weight}
\label{sec:diagonal_weight}
The discussion above informs us that if we use the diagonal part of the Wigner matrices $D^L_{MM}$ as the weight function in the rotational average, then the \emph{orientation dependence of the anisotropies} is retained, i.e.,
\begin{align}    \label{I_LM}
    I^{L}_{M}(\uvec{n}_{1}, \uvec{n}_{2})
    = \int \d\mu(\rot)
    D^{L*}_{M M}(\rot)
    C(\rot\uvec{n}_{1}, \rot \uvec{n}_{2}).
\end{align}
These rotational averages $I^{L}_{M}(\un_1, \un_2)$ are sensitive to the azimuthal index $M$. 
These related to the BipoSH coefficients as
\begin{align}
\label{eq:ILM}
    I^{L}_{M}(\uvec{n}_{1}, \uvec{n}_{2})
    = \frac{1}{2L+1} 
    \sum_{\ell_{1}\ell_{2}}
    A^{L M}_{\ell_{1}\ell_{2}}
    \bi^{L M}_{\ell_{1}\ell_{2}}(\uvec{n}_{1}, \uvec{n}_{2}).
\end{align}
Hence $I^L = \sum_M I^L_M$, i.e., in contrast to the character weight, using the diagonal Wigner matrices retains the azimuthal dependence by skipping the sum over azimuthal indices.%avoids summing over the azimuthal indices, thus retaining the azimuthal dependence.%

Let us now explicitly see the behaviour of $I^{L}_{M}$ under statistical isotropy. For an SI field, the two-point correlation function is invariant,
$C(\rot \un_1, \rot \un_2) = C(\un_1, \un_2)$ for every $\rot \in SO(3)$, so it factors out of
the integrand in Eq.~\eqref{I_LM} and
\begin{align}
\begin{split}
\label{eq:ILM_SI_real}
I^{L}_{M}(\un_1, \un_2)
=& \; C(\un_1, \un_2) \int \d\mu(\rot)\, D^{L*}_{MM}(\rot)\\
=& \; C(\un_1, \un_2)\, \delta_{L0}\, \delta_{M0},
\end{split}
\end{align}
where we have used $D^{0}_{00}(\rot) = 1$ together with the orthogonality relation
Eq.~\eqref{eq:wigner_ortho}. 

The same conclusion follows in the harmonic route: SI fixes the harmonic two-point function to
$\Braket{a_{\ell_1 m_1} a_{\ell_2 m_2}} = (-1)^{m_1} C_{\ell_1} \delta_{\ell_1 \ell_2}
\delta_{m_1 -m_2}$, so that the BipoSH coefficients reduce to
\begin{align}
\label{eq:A_SI}
A^{LM}_{\ell_1 \ell_2}
= (-1)^{\ell_1} \sqrt{2\ell_1 + 1}\, C_{\ell_1}\, \delta_{\ell_1 \ell_2}\,
\delta_{L0}\, \delta_{M0},
\end{align}
and, substituting in Eq.~\eqref{eq:ILM},
\begin{align}
\label{eq:ILM_SI}
I^{L}_{M}(\un_1, \un_2)
= \delta_{L0}\, \delta_{M0} \sum_{\ell} \frac{2\ell+1}{4\pi} C_{\ell}\,
P_{\ell}(\un_1 \cdot \un_2).
\end{align}
The surviving $L = 0$ component is thus the isotropic angular correlation function itself,
consistent with Eq.~\eqref{eq:ILM_SI_real}. Note that Eq.~\eqref{eq:ILM_SI} is equivalent to performing the rotational average with uniform weight (Eq.~\eqref{eq:I_uni}), and this connection is straightforward --- the uniform weight average retains only the $SO(3)$ invariant component of the two-point function. 
A signal at $L > 0$ in $I^{L}_{M}$ is thus a signature of statistical isotropy
violation in the two-point function.

Two points are worth emphasizing here. The first is what $I^{L}_{M}$ at $L > 0$ actually probes.
For a Gaussian field, the two-point function provides a complete statistical description, so the
vanishing of $I^{L}_{M}$ for all $L > 0$ is equivalent to the field being SI. For a non-Gaussian
field, $I^{L}_{M}$ tests only the two-point function; the three- and higher-point functions may independently
violate rotational invariance, so the field may still be nSI even when its two-point function is isotropic (refer Appendix \ref{appendix}). %we shall see this for three-point function in  %Sec.~\ref{sec:rotavg_3pt}..

Second, a statistical
inhomogeneity of the underlying three-dimensional perturbations will generically appear as an
anisotropy to an off-centre observer. 
The rotational averages test the observed sky for departures from SI. Since the CMB temperature is a field on $S^2$, a statistic constructed from $\Delta T(\un)$ alone cannot discern the origin of such departures, which may arise from either statistical inhomogeneity or statistical anisotropy in the underlying three-dimensional primordial perturbations.

Similar to the case of the character weight, one can further take the direction averages of the quantity $\abs{I^{L}_M(\un_1,\un_2)}^2$ to arrive at a positive definite global measure of nSI components of $C(\un_1, \un_2)$:
\begin{align} \label{relation}
    \kappa^L_M
    \equiv \int \frac{\d \Omega_{\un_2}}{4 \pi}
    \int \frac{\d \Omega_{\un_1}}{4 \pi}
    \abs{I^{L}_M(\un_1,\un_2)}^2.
\end{align}
In the harmonic space, this measure takes the form
\begin{align}  \label{kappa}
   \kappa^L_M
   = \frac{1}{(4 \pi)^2(2L+1)^{2}}
     \sum_{\ell_1 \ell_2}
    \abs{A^{LM}_{\ell_{1}\ell_{2}}}^{2},
\end{align}
Interestingly, if we project the rotational average $I^{L}_{M}(\uvec{n}_{1}, \uvec{n}_{2})$ on the BipoSH basis, we recover the BipoSH coefficients:
\begin{align}
\begin{split}
    A^{L M}_{\ell_{1}\ell_{2}}
    =& 
   (2L+1)\\
   &\times \int \d \Omega_{\uvec{n}_1} \int \d \Omega_{\uvec{n}_2}\
   I^{L}_{M}(\uvec{n}_{1}, \uvec{n}_{2})
   \bi^{L M*}_{\ell_{1}\ell_{2}}(\uvec n_1,\uvec n_2).
   \end{split}
\end{align}
Also, summing over the internal ranks in the above expression, one can represent the reduced BipoSH coefficients (introduced in~\cite{Hajian:2006ud}) in terms of the rotational averages:
\begin{align}
\begin{split}
    A^{L M}
    =& \sum_{\ell_1 \ell_2} A^{L M}_{\ell_{1}\ell_{2}}\\
    =&(2L+1)\\
   &\times \int \d \Omega_{\uvec{n}_1} \int \d \Omega_{\uvec{n}_2}
   I^{L}_{M}(\uvec{n}_{1}, \uvec{n}_{2})
   \sum_{\ell_1 \ell_2}
   \bi^{L M*}_{\ell_{1}\ell_{2}}(\uvec n_1,\uvec n_2).
   \end{split}
\end{align}
In the next section, we present an example to demonstrate explicitly that a non-vanishing weighted rotational average of the correlation function indicates the presence of an anisotropic signal.

\section{Illustrative example}
{We consider the phenomenological model of dipole modulation, widely adopted in the Planck analyses~(\cite{Planck_2013,Planck:2015igc,Planck_2018}) to characterize the Cosmic Hemispherical Asymmetry~(\cite{Shaikh_2019}) observed in CMB}: 
\begin{align}   \label{dm_model}
\Delta T(\un) 
&= \Delta T_{\rm iso}(\un) [ 1 + \epsilon (\un \cdot \uvec{p})],
\end{align}
where $\uvec{p}$ defines a special direction on the sky and $\epsilon$ is a smallness parameter. 
The real space angular correlation function takes the form
\begin{align}
\begin{split}   
\label{eq:C_dipole_mod}
C(\un_1, \un_2) 
=& C_{\rm iso}(\un_1 \cdot \un_2) 
  [ 1 + \epsilon (\un_1 \cdot \uvec{p} + \un_2 \cdot \uvec{p}) \\
  &+ \epsilon^2 (\un_1 \cdot \uvec{p})(\un_2 \cdot \uvec{p}) ].
\end{split}
\end{align}
The leading term $C_{\rm iso}$ is the isotropic background. The $\mathcal{O}(\epsilon)$ terms represent a dipolar modulation along the special direction. The subdominant $\mathcal{O}(\epsilon^2)$ term captures the dipole-dipole correlation inducing a quadrupole-like modulation.

As the isotropic factor $C_{\rm iso}$ remains untouched by the $SO(3)$ integration Eq.~\eqref{I_LM}, the rotational averages are significantly simpler to compute than their harmonic counterparts $A^{LM}_{\ell_1 \ell_2}$.  These quantities take the following forms for the present example: 
\begin{align}    
\label{ilm}
I^L_M(\un_1, \un_2) 
&=  C_{\rm iso}(\un_1 \cdot \un_2)
\Big\{\delta_{L0} \delta_{M0} \nonumber \\
&\quad + 
\epsilon
\delta_{L1} \frac{4\pi}{9} Y_{1M}^*(\uvec{p}) 
\left[ Y_{1M}(\un_1) + Y_{1M}(\un_2) \right] \nonumber \\
&\quad + \epsilon^2 
\frac{8\pi^{3/2}}{3(2L+1)^{3/2}}
\cg_{1 0 1 0}^{L 0}
Y_{LM}^*(\uvec{p}) \nonumber\\
&  \left. \quad \times \bi^{LM}_{1 1}(\un_1,\un_2) \right\} .
\end{align}
$I^L_M(\un_1, \un_2)$ extracts the multipole structure of the correlation function, explicitly mapping how the special direction $\uvec{p}$ couples to the line-of-sight directions ($\un_1$, $\un_2$) to generate the nSI features.
The monopole ($I^0_0$) captures the dominant rotationally invariant background, it also gets an $\mathcal{O}(\epsilon^2)$ contribution from the modulation factor. The dipole ($I^1_M$) arises at $\mathcal{O}(\epsilon)$, linearly coupling the special direction to each line-of-sight independently. For the $\mathcal{O}(\epsilon^2)$ term, the Clebsch-Gordan coefficient $\cg^{L0}_{1010}$ restricts the contributions to the $L=0$ and $L=2$ sectors. The $\mathcal{O}(\epsilon^2)$ quadrupole ($I^2_M$) contribution comes from the dipole-dipole cross-correlation, characterized by the coupling of the BipoSH basis $\bi^{2M}_{11}(\un_1, \un_2)$ with the special direction.

The geometric clarity of the forms of $I^L_M$ are especially evident when compared to the BipoSH coefficients:
\begin{align}
\begin{split}
A^{LM}_{\ell_1 \ell_2} 
&= (-1)^{\ell_1} \sqrt{2\ell_1 + 1} C_{\ell_1} \delta_{\ell_1 \ell_2} \delta_{L0} \delta_{M0} \nonumber \\
&\quad + \epsilon
\delta_{L1} \frac{2\sqrt{\pi}}{3} Y_{1M}^*(\uvec{p})
\sqrt{\frac{(2\ell_1+1)(2\ell_2+1)}{3}}\\
&\quad\times \cg_{\ell_1 0 \ell_2 0}^{1 0} 
\left( C_{\ell_1} + C_{\ell_2} \right)\nonumber \\
&\quad + \epsilon^2
\sqrt{\frac{4\pi}{2L+1}} 
\cg_{1 0 1 0}^{L 0} Y_{LM}^*(\uvec{p})\\
&\quad \times \sum_{\ell} (2\ell+1) C_{\ell} (-1)^{\ell_1 - 1}
\begin{Bmatrix}
\ell_1 & \ell_2 & L \\ 1 & 1 & \ell
\end{Bmatrix}
\cg_{\ell 0 1 0}^{\ell_1 0} \cg_{\ell 0 1 0}^{\ell_2 0}.
\end{split}
\end{align}
Although rotational averages fully encode the breaking of rotational symmetries, their explicit line-of-sight dependence may make them challenging to analyze and represent in the context of complex phenomenological models or real CMB data. Instead, the quantity $\kappa^L_M$ provides a more practical measure: it averages out the explicit line-of-sight dependence, but still retains information of the orientation of the special direction through its azimuthal index $M$.
For the present model, these are
\begin{align}   \label{kappa_th}
\kappa^0_0 
&= \frac{1}{(4\pi)^2} 
\sum_{\ell} \left[ (2\ell+1) C_{\ell}^2 + \frac{4\epsilon^2}{3} 
(\ell+1) C_{\ell} C_{\ell+1} \right], \\
\kappa^1_M 
&= \frac{\epsilon^2}{162\pi} | Y_{1M}(\uvec{p}) |^2 
\sum_{\ell} (\ell+1) \left( C_{\ell} + C_{\ell+1} \right)^2,\\
\kappa^2_M
&=  \mathcal{O}(\epsilon^4).
\end{align}
The dominant contribution to the monopole $\kappa^0_0$ stems from the rotationally invariant unmodulated  background. The monopole also receives a positive $\mathcal{O}(\epsilon^2)$ shift from the modulation factor. Because the monopole lacks azimuthal dependence, this shift is independent of the orientation of the special direction. Crucially, \emph{the dipole contribution $\kappa^1_M \propto \epsilon^2 |Y_{1M}(\uvec{p})|^2$ probes the geometry of the special direction}. For example, a dominant $M=0$ projection implies the special axis is aligned with the $z$-axis of the coordinate system; while dominant $M=\pm 1$ projections implies the special direction lies in the equatorial plane. Finally, the quadrupole power $\kappa^2_M$ gets its leading contribution of order $\mathcal{O}(\epsilon^4)$ through the dipole-dipole interaction term.

\section{Numerical implementation}
To establish the implementability of nSI measurements in the real space approach, we numerically compute the quantities in Eqs.~\eqref{I_LM} and~\eqref{kappa}, and compare them with the traditional harmonic space results. To compute $I^L_M(\un_1, \un_2)$ numerically, we define its estimator as
\begin{align}  \label{I_L_est}
    \tilde{I}^{L}_{M}(\un_1,\un_2)
    = 
    \frac{1}{N_R}\sum_{i}^{N_R} D_{MM}^{L*}(\rot_i)\Delta T(\rot_i \un_1)\Delta T(\rot_i \un_2),
\end{align}
where $N_R$ is the number of rotations sampled.
The sampling in the above equation is to be performed with respect to the Haar measure on $SO(3)$. To achieve this, we use the fact that $SU(2)$ is the double cover of $SO(3)$. Since the group of unit quaternions is isomorphic to 
$SU(2)$, we uniformly generate unit quaternions using Shoemake's formula~(\cite{shoemake}). Shoemake's formula parametrizes the three-sphere $S^3$ --- which is the group manifold of $SU(2)$ --- using two angular variables and one radial parameter, effectively viewing the three-sphere as two coupled circles. This then reduces the problem to uniformly sampling points on these coupled circles, with the coordinates of the points forming the components of the unit quaternion $(q)$. 
Finally, the unit quaternion is mapped to the corresponding rotation in $SO(3)$, with both $q$ and $-q$ identifying the same rotation. Further discussions on sampling on $SO(3)$ can be found in~(\cite{shoemake}). 
\begin{figure}  
    \centering
    \includegraphics[width=1.0\linewidth]{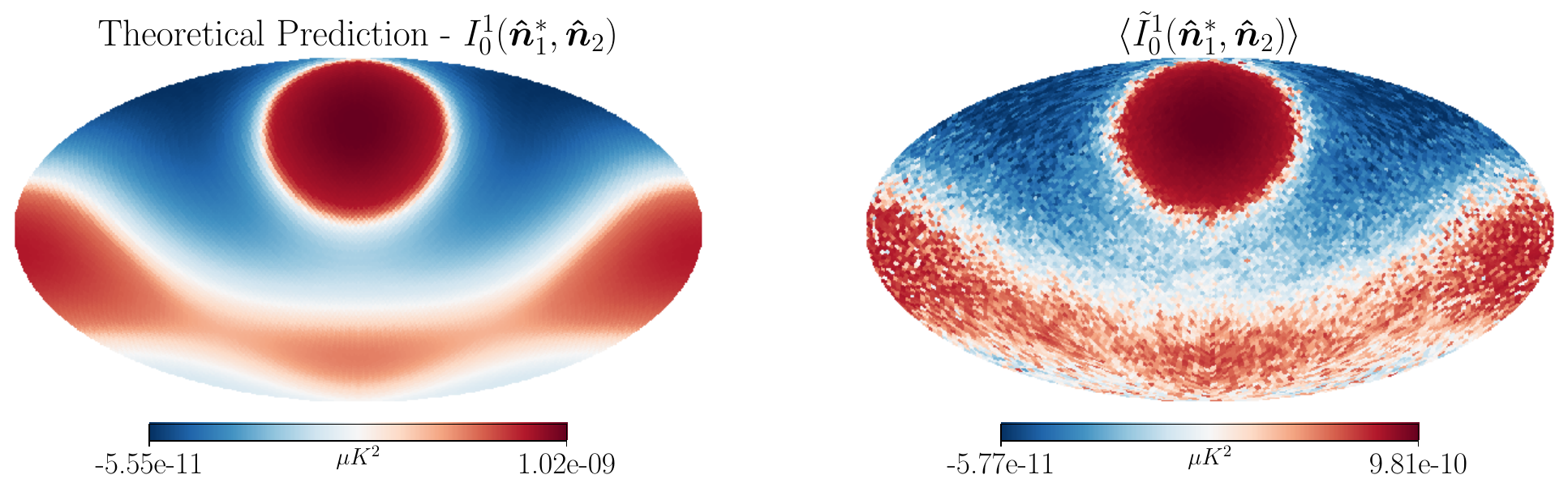}
    \caption{\emph{Left:} Theoretical prediction of $I_0^1(\un_1^*,\un_2)$ plotted for a dipole modulation map ($\mathtt{nside} = 32$) with axis $\uvec{p} = \un_z$ and modulation amplitude $\epsilon = 0.5$ (cf.~Eq.~\eqref{ilm}). $\un_1^*$ indicates a fixed line-of-sight direction, in this case, $\un_1^*(\theta,\phi) = (\pi/4,0)$. \emph{Right:} Ensemble averaged $\tilde{I}_0^1(\un_1^*,\un_2)$ over 100 realizations with the same dipole modulation parameters.}
     \label{I1_comp}
\end{figure}

%Figure~\ref{I1_comp} shows the computed ensemble averaged $\tilde{I}^{L}_{M}(\un_1,\un_2)$ using Eq.~\eqref{I_L} for the dipole modulation map with one line-of-sight point fixed ($\un_1^*$) over all the 100 realizations. Figure~\ref{I00} shows the estimation of the isotropic component of the two-point function and the variance of the estimator $\tilde{I}^0$, exhibiting a close agreement with the cosmic variance.

\begin{figure}  
    \centering
    \includegraphics[width=1.0\linewidth]{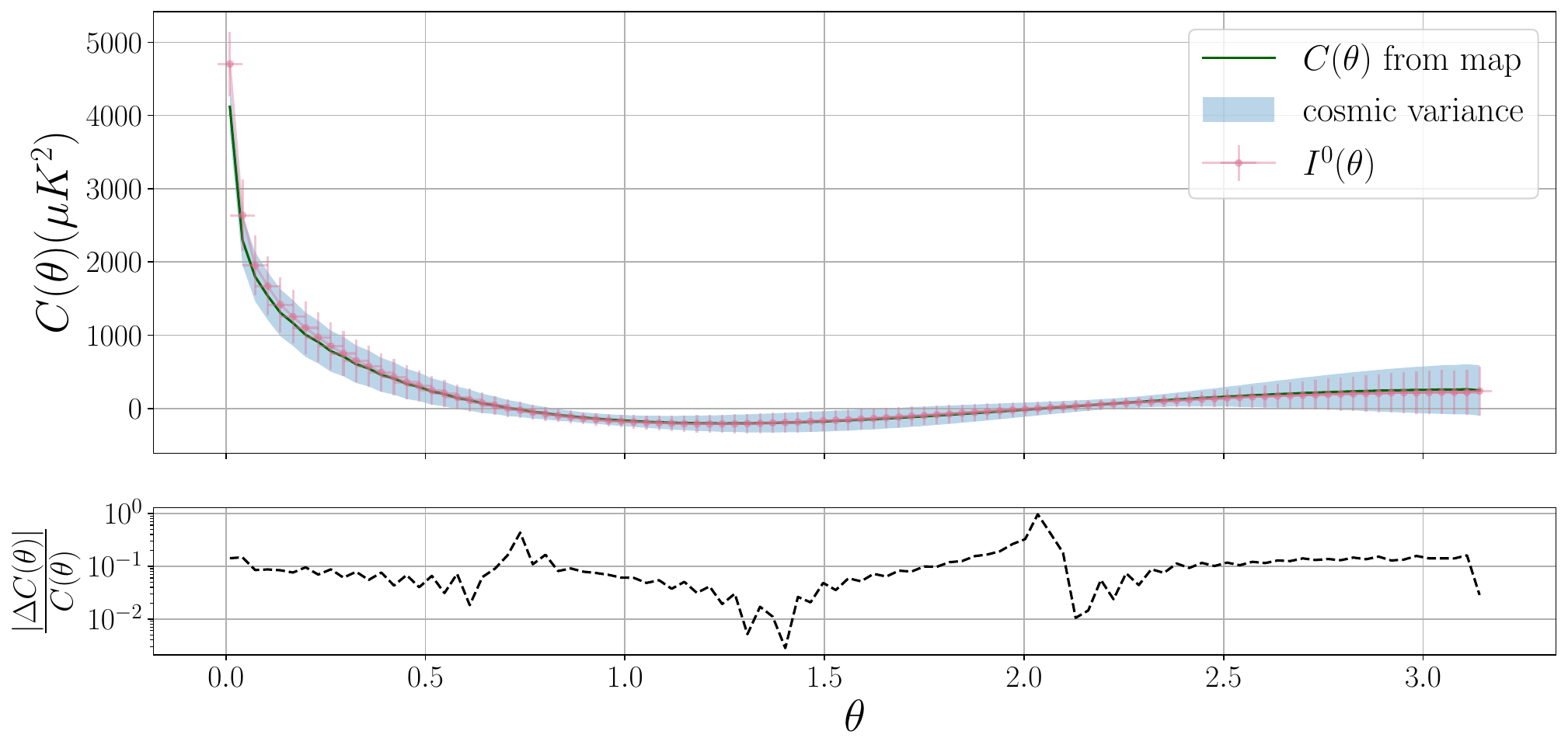}
    \caption{\emph{Top panel}: The statistically isotropic component of the two-point correlation function estimated using $\chi^L$ weighted rotational average (cf.~Eq.~\eqref{I_L}) over 1000 realizations of a dipole modulated map ($\mathtt{nside} = 32$) with $\epsilon=0.5$. The markers indicate the mean of the 100 bins considered, horizontal and the vertical error bars indicate the bin width and the variance of the estimator $\tilde{I}^0$, respectively. The shaded region is the analytical cosmic variance, while the solid line is the theoretical $C(\theta)$. \emph{Bottom panel}: The fractional difference between the estimated two-point correlation function and the theoretical $C(\theta)$. The two peaks observed in the fractional difference are artifacts of the zero crossing of $C(\theta)$.}
    \label{I00}
\end{figure}

We use the \texttt{HEALPix}\footnote{\url{https://healpix.sourceforge.io}} package~(\cite{Gorski_2005}) for map computations and visualizations. Figures~\ref{I1_comp} and \ref{I00} show the ensemble averaged estimates of  $\tilde{I}^{1}_{M}(\un_1,\un_2)$ and $\tilde{I}^0(\theta)$ (Eq.~\eqref{I_L_est}) for the dipole modulation model. In Figure~\ref{I1_comp}, one line-of-sight direction ($\un_1^*$) is fixed over all realizations. It can be seen that the map of $\left<\tilde{I}_0^1(\un_1^*,\un_2)\right>$ closely matches the theoretical result. Figure~\ref{I00} shows that the real space estimator retrieves the isotropic component of the two-point function, while the variance of the estimator closely tracks the analytical cosmic variance.

Similarly, the estimator for $\kappa_M^L$ can be expressed as, 
\begin{align}  \label{kappa_est}
    \tilde{\kappa}^{L}_{M}
    = 
    \frac{1}{n_{p}}\sum_{i}^{n_{p}} \tilde{I}^{L}_{M}[(\un_1,\un_2)_i]
    \tilde{I}^{L*}_{M}[(\un_1,\un_2)_i].
\end{align}
The estimator in Eq.~\eqref{kappa_est} is computed by uniformly sampling the sphere and obtaining $n_{p}$ pairs of line-of-sight points. 

Note that while the estimator $\tilde{I}^{L}_{M}(\un_1, \un_2)$ is unbiased, the estimator
$\tilde{\kappa}^{L}_{M}$ carries a bias $B^{L}_{M} = \Braket{\tilde{\kappa}^{L}_{M}} -
\kappa^{L}_{M}$, owing to its non-linear dependence on $\tilde{I}^{L}_{M}(\un_1, \un_2)$.
In the harmonic space it takes the form:
\begin{align}
\begin{split}
\label{eq:bias}
B^{L}_{M}
=& \frac{1}{(4\pi)^2 (2L+1)^2}
\sum_{\ell_1 \ell_2} \sum_{m_1 m_2 m_1' m_2'}\\
&\times \cg^{LM}_{\ell_1 m_1 \ell_2 m_2}\,
\cg^{LM}_{\ell_1 m_1' \ell_2 m_2'}\\
&\times \Big[
\Braket{a_{\ell_1 m_1} a_{\ell_2 m_2} a^{*}_{\ell_1 m_1'} a^{*}_{\ell_2 m_2'}}\\
&\qquad - \Big\langle a_{\ell_1 m_1} a_{\ell_2 m_2} \Big\rangle
\Braket{a^{*}_{\ell_1 m_1'} a^{*}_{\ell_2 m_2'}}
\Big],
\end{split}
\end{align}
The bias contains the four-point functions and pairs of two-point functions. %Both the terms contribute to $B^{L}_{M}$ at $L \neq 0$ even for a statistically isotropic sky.
Even for a statistically isotropic, Gaussian sky, $B^{L}_{M}$ at $L \neq 0$ is non-vanishing due to the residual contribution from the four-point function that survives subtraction of the second term. 
Consequently, a non-zero $\tilde{\kappa}^{L}_{M}$ at
$L \neq 0$ is not by itself evidence of broken isotropy, and must be assessed against the expected bias. This is a caveat of the quadratic nature of $\kappa^L_M$, in contrast, the rotational average estimator $\tilde{I}^{L}_{M}$ is unbiased and hence free of such caveat.

As the bias $B^{L}_{M}$ depends on both nSI and non-Gaussian features of the field, computing it in general requires modelling these features.
However, under the assumption of Gaussianity and SI, the bias and can be obtained analytically~(\cite{Hajian:2003qq}) as:
\begin{align}
\begin{split}
\label{eq:bias_SI}
B^{L}_{M}
= \frac{1}{(4\pi)^2 (2L+1)}
\Big[ &\sum_{\ell_1} \sum_{\ell_2 = \abs{L-\ell_1}}^{L+\ell_1}
C_{\ell_1} C_{\ell_2}\\
+ &\sum_{\ell_1} \sum_{\ell_2 = \abs{L-\ell_1}}^{L+\ell_1}
(-1)^{L} C_{\ell_1} C_{\ell_2} \delta_{\ell_1 \ell_2} \Big].
\end{split}
\end{align}

% Note that while the estimator $\tilde{I}^{L}_{M}(\un_1,\un_2)$ is unbiased, the estimator $\tilde{\kappa}^{L}_{M}$ has a bias $B_M^L =  \Braket{\tilde{\kappa}_M^{L}} - \kappa_M^L$..
% due to its non-linear dependence on $\tilde{I}^{L}_{M}(\un_1,\un_2)$.
% % \begin{equation}
% %     B_M^L =  \Braket{\tilde{\kappa}_M^{L}} - \kappa_M^L.
% % \end{equation}
% Deriving an analytical expression for the bias $B_M^L$ is non-trivial in the presence of nSI. However, under the assumption of Gaussianity and SI, the bias can be computed following~(\cite{Hajian:2003qq}).
% In our numerical implementation, we focus on comparing the harmonic   and real space results for $\kappa^L_M$; therefore, we evaluate the biased expressions for both cases.

\begin{figure}
    \centering
    \includegraphics[width=1.0\linewidth]{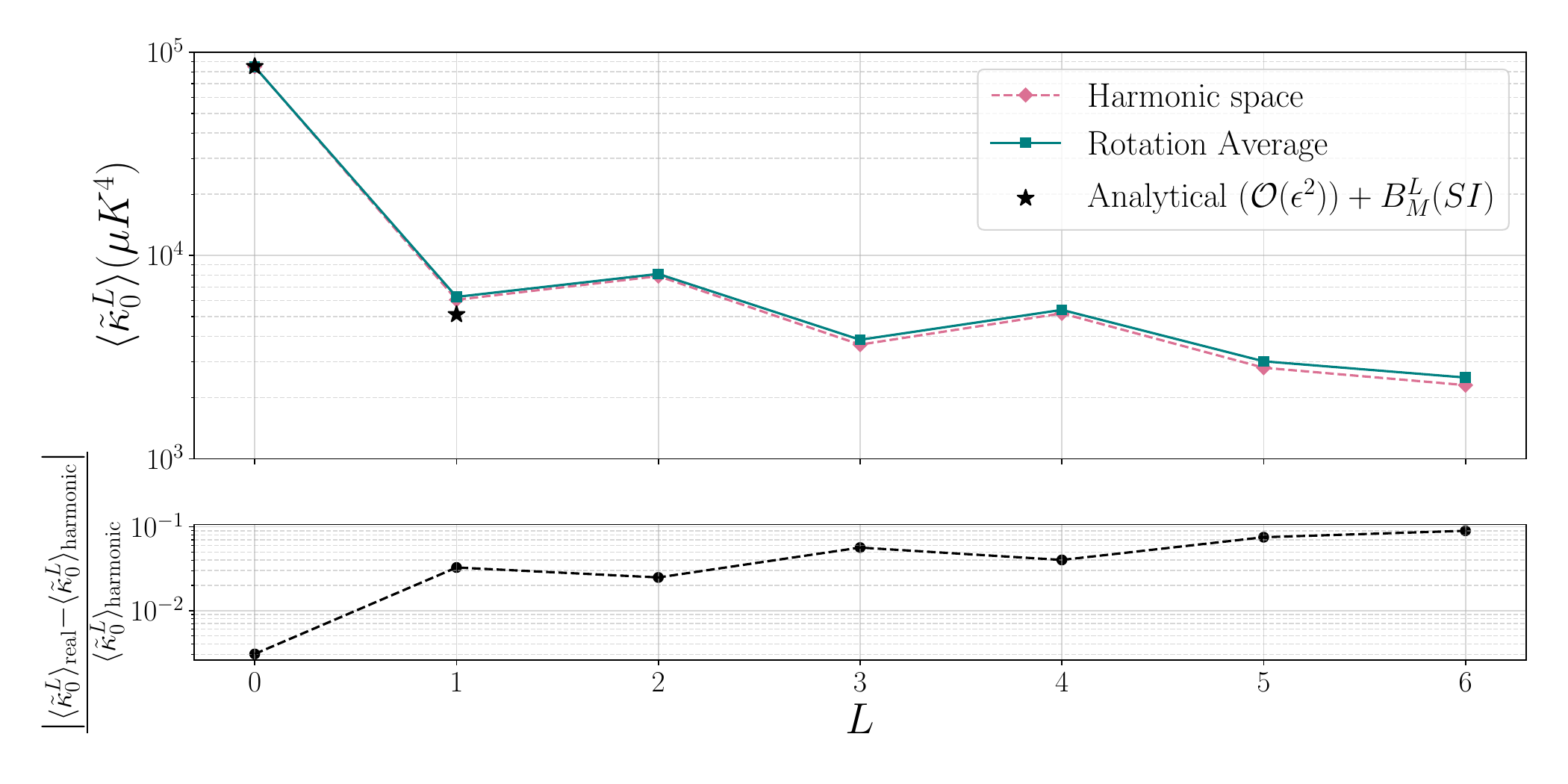}
    \caption{\emph{Top panel}: $\Braket{\tilde{\kappa}_0^L}$ from rotational averages (solid green line) and the corresponding harmonic space estimator (dashed red line), ensemble averaged over 1000 realizations of the dipole modulated map ($\mathtt{nside} = 32$) with amplitude $\epsilon=0.5$ and axis $\uvec{p} = \un_z$. The black markers indicate the analytically computed $\kappa^L_0$ up to $\mathcal{O}(\epsilon^2)$ for $L \leq 1$ with the SI component of the bias added. \emph{Bottom panel}: The fractional difference between the real  and harmonic space estimators. The number of rotations sampled are of the order of $10^5$ for 5000 line-of-sight pairs.}
    \label{kappa_comp}
\end{figure}

In Fig.~\ref{kappa_comp}, we compare the estimators for $\tilde{\kappa}_0^L$ computed from rotational averages and harmonic space, along with theoretical estimations, evaluated up to $\mathcal{O}(\epsilon^2)$ for $L \leq 1$.
To ensure a better comparison, we have added the analytical Gaussian and SI bias of
Eq.~\eqref{eq:bias_SI} to the theoretical estimations
(see~\cite{Hajian:2003qq} for further details on the bias).

% To ensure a better comparison, we have added the analytical SI bias for $\tilde{\kappa}^L_0$,
% \begin{align}
% \begin{split}
%     B^L_0 
%     = \frac{1}{(4 \pi)^2 (2 L + 1)} \sum_{\ell_1} 
%     \sum_{\ell_2 = \abs{L-\ell_1}}^{L + \ell_1}
%     C_{\ell_1} C_{\ell_2} + \\
%     \sum_{\ell_1} 
%     \sum_{\ell_2 = \abs{L-\ell_1}}^{L + \ell_1}
%     C_{\ell_1} C_{\ell_2}(-1)^L \delta_{\ell_1 \ell_2}
%     \end{split}
% \end{align}
% with the theoretical estimations (see~\cite{Hajian:2003qq} for further details on the bias).

Importantly, the real  and harmonic space results are in agreement up to a fractional deviation of $\leq 10^{-1}$ for low multipoles $L \leq 6$. 
Note that, the bias-added theoretical prediction (Eq.~\eqref{kappa_th}) falls below both the real  and harmonic space estimators, particularly at the $L=1$ scale. This discrepancy can be attributed to contributions from the $\mathcal{O}(\epsilon^3)$ terms, as well as the nSI component of the bias which are not taken into account.

We further note that the fractional difference between the real  and harmonic space results grow with increasing $L$. This deviation stems from the limitation of discretely sampling the rotation group.
The functional behavior of the $SO(3)$ integrand (Eq.~\eqref{I_L}) is dictated by the Wigner matrices; specifically, the character function of rotation is proportional to  $\sin(\omega(L+1/2))$, where $\omega$ is the rotation angle~(\cite{Varshalovich:1988ifq}). 
As a result, the integrand becomes highly oscillatory at higher multipoles. Consequently, for a fixed number of sampled rotations, the discrete integration becomes less accurate at higher multipoles. Such sampling errors might be rectified, keeping the computation cost in mind, by suitable scaling the number of samples with $L$.

\section{Discussion}
The real-space formalism shows promising agreement to the harmonic space counterpart, up to $\mathcal{O}(10^{-1})$ fractional deviation in the measure $\kappa^L_M$, which can be further improved with appropriate scaling of the number of rotations sampled at higher $L$'s.
Apart from being an alternative route to extract nSI signatures, the real-space construction is
also a promising route to mitigate the non-trivial partial sky effects such as mode coupling and
leakage between different polarization modes (see, e.g.,~\cite{Hivon:2001jp, Lewis:2001hp}) that
arise in the harmonic space decomposition.
A partial sky analysis will be pursued in a future work.

% Apart from being an alternative route to extract nSI signatures, the estimator is also useful in mitigating the non-trivial partial sky effects such as mode coupling and leakage between different polarization modes (see, e.g.,~\cite{Hivon:2001jp, Lewis:2001hp}) that arise in the harmonic space decomposition.  

The rotational averages extend to higher-order correlation functions. We have shown
this explicitly for the three-point correlation function (Appendix \ref{appendix}), which can be induced if the field is non-Gaussian. If the three-point function is
statistically isotropic, $I^{(3)L}_{M}$ is supported entirely at $L = 0$, regardless of the shape
and amplitude of the bispectrum; for a zero-mean Gaussian field, $I^{(3)L}_{M}$ vanishes
identically for every $L$, including $L = 0$.

% The rotational averages can readily be extended to higher order correlation functions; for example,
% {\color{red} a general $N$-point correlation function on a sphere $\mathcal{T}(\un_1, \un_2,..., \un_n) = \Braket{\Delta T(\un_1) \Delta T(\un_2),..., \Delta T(\un_n)}$ can be expanded in the N-polar spherical harmonics,
% \begin{align}
% \mathcal{N}_{\boldsymbol{l}\boldsymbol{s},\tau}^{LM}(\hat{\mathbf{n}}) = \sum_{\boldsymbol{m}} \Gamma_{\boldsymbol{l}\boldsymbol{m}\boldsymbol{s}}^{LM}(\tau)\, Y_{l_1}^{m_1}(\hat{\mathbf{n}}_1) \cdots Y_{l_\nu}^{m_\nu}(\hat{\mathbf{n}}_\nu),
% \end{align}
% where $\Gamma_{\boldsymbol{l}\boldsymbol{m}\boldsymbol{s}}^{LM}(\tau)$ is the coupling tensor. 
% The N-poSH behaves analogously to the bipolar spherical harmonics under rotation and hence the rotational symmetry breaking in $\mathcal{T}(\un_1, \un_2,...., \un_n)$ can be captured in the quantity
% \begin{align}
%     I^{(N)L}_{M}(\un_1, \un_2,..., \un_n)
%     =& \int \d\mu(\rot)
%     D^{L*}_{M M}(\rot)\\ \nonumber
%     &\times \mathcal{T}(\rot \un_1, \rot \un_2,...., \rot \un_n).
% \end{align} 
% For $N=3$, which is the 3-point function, the rotational average measure can potentially probe isotropy violation signatures encoded in primordial non-Gaussianity~(\cite{Kamionkowski_2011})}.

% 
Similarly this framework can be used to study the weak gravitational lensing of CMB~(\cite{Carvalho:2010rz, Bucher:2010iv}) and can also be generalized to higher spin fields such as CMB polarization. Since this framework is explicitly rooted in symmetry, it can be used to study phenomena with known breakdowns of rotational symmetry, for example, in models of non-trivial topologies~(\cite{Lachieze-Rey:1995qrb, Hajian:2003qq, Bond_1998}).
 Furthermore, this geometric framework can be extended to study isotropy violations across a broader class of fields including astrophysics, nuclear physics~(\cite{Friar:1981zza, Ershov:2004va}), and atomic and molecular physics~(\cite{manakov_multipole_2002, meremianin_three-body_2005}).

\begin{acknowledgments}
The authors acknowledge the HPC facilities at RRI. 
VR thanks Rishav Saha for useful discussions. 
DM thanks Harkirat Singh Sahota for valuable comments and suggestions. 
DM thanks Raman Research Institute for support through postdoctoral fellowship. 
TS acknowledges support from the J.~C.~Bose Grant of Anusandhan National Research Foundation, India.
\end{acknowledgments}

\appendix 
\section{Rotational averages of CMB three-point correlation function}
\label{appendix}
The rotational average formalism can be extended to N-point function as, 
\begin{align}
\begin{split}
\label{eq:IN_LM}
I^{(N)L}_{M}(\un_1, \un_2,..., \un_N)
=& \int \d\mu(\rot)\, D^{L*}_{MM}(\rot)
 \mathcal{N}(\rot \un_1, \rot \un_2, ...,\rot \un_N),
\end{split}
\end{align}
with $\mathcal{N}(\rot \un_1, \rot \un_2, ...,\rot \un_N) = \Braket{\Delta T(\un_1) \Delta T(\un_2).... \Delta T(\un_N)}$, a real function defined on $\underbrace{S^2 \times S^2 \times \cdots \times S^2}_{N \text{ times}}$.

 We focus specifically on the three-point function here.
We denote the three-point function as $\mathcal{T}(\un_1, \un_2, \un_3) = \Braket{\Delta T(\un_1) \Delta T(\un_2) \Delta T(\un_3)}$, a real function defined on $S^2 \times S^2 \times S^2$.
The harmonic representation of a three-point function is provided by the tripolar spherical harmonic (TripoSH) basis, which couples the three internal ranks ($\ell_1, \ell_2, \ell_3$) corresponding to the three directions ($\un_1, \un_2, \un_3$) into a state of total angular momentum $L$ through an intermediate rank $\ell_{23}$~(\cite{Varshalovich:1988ifq}),
\begin{align}
\begin{split}
\label{eq:triposh}
\tri_{\ell_1 \ell_2 \ell_3 \ell_{23}}^{LM}(\un_1, \un_2, \un_3) 
\equiv &\{Y_{\ell_1}(\un_1) \otimes \{Y_{\ell_2}(\un_2) \otimes Y_{\ell_3}(\un_3)\}_{\ell_{23}}\}_{LM}\\
=& \sum_{m_1 m_2 m_3 m_{23}}
\cg_{\ell_1 m_1 \ell_{23} m_{23}}^{LM}\,
\cg_{\ell_2 m_2 \ell_3 m_3}^{\ell_{23} m_{23}} Y_{\ell_1 m_1}(\un_1) Y_{\ell_2 m_2}(\un_2) Y_{\ell_3 m_3}(\un_3),
\end{split}
\end{align}
where the triangle inequalities
$|\ell_2 - \ell_3| \le \ell_{23} \le \ell_2 + \ell_3$ and
$|\ell_1 - \ell_{23}| \le L \le \ell_1 + \ell_{23}$, together with the azimuthal conditions
$m_{23} = m_2 + m_3$ and $M = m_1 + m_2 + m_3$ are assumed.
The three-point function is then expanded as
\begin{align}
\label{eq:3pcf_triposh}
\mathcal{T}(\un_1, \un_2, \un_3)
=
\sum_{LM} \sum_{\ell_1 \ell_2 \ell_3 \ell_{23}}
t_{\ell_1 \ell_2 \ell_3 \ell_{23}}^{LM}
\tri_{\ell_1 \ell_2 \ell_3 \ell_{23}}^{LM}(\un_1, \un_2, \un_3),
\end{align}
where $t_{\ell_1 \ell_2 \ell_3 \ell_{23}}^{LM}$ are the TripoSH coefficients.

The essential property for our purpose is that the TripoSH basis, like the BipoSH basis,
transforms under a rotation
$\rot \in SO(3)$ analogously to Eq.~\eqref{eq:biposh_rot}~(\cite{Varshalovich:1988ifq}):
\begin{align}
  \begin{split}
\label{eq:triposh_rot}
&\tri_{\ell_1 \ell_2 \ell_3 \ell_{23}}^{LM}(\rot \un_1, \rot \un_2, \rot \un_3)
= \sum_{M'} D^{L}_{M' M}(\rot)\,
\tri_{\ell_1 \ell_2 \ell_3 \ell_{23}}^{LM'}(\un_1, \un_2, \un_3).
  \end{split}
\end{align}
A rotation via the Wigner-$D$ matrix thus mixes only the azimuthal index $M$, leaving the
total rank $L$ and the internal ranks
$\ell_1, \ell_2, \ell_3, \ell_{23}$ untouched. Adopting the diagonal
Wigner elements as the weight, as in Sec.~\ref{sec:diagonal_weight},
the rotational average of the three-point function is
\begin{align}
\begin{split}
\label{eq:I3_LM}
I^{(3)L}_{M}(\un_1, \un_2, \un_3)
=& \int \d\mu(\rot)\, D^{L*}_{MM}(\rot)
 \mathcal{T}(\rot \un_1, \rot \un_2, \rot \un_3),
\end{split}
\end{align}
and substituting Eqs.~\eqref{eq:3pcf_triposh}
and~\eqref{eq:triposh_rot} into the above definition, one obtains
\begin{align}
\begin{split}
\label{eq:I3_t}
I^{(3)L}_{M}(\un_1, \un_2, \un_3)
=& \frac{1}{2L+1} \sum_{\ell_1 \ell_2 \ell_3 \ell_{23}}
t_{\ell_1 \ell_2 \ell_3 \ell_{23}}^{LM} \tri_{\ell_1 \ell_2 \ell_3 \ell_{23}}^{LM}(\un_1, \un_2, \un_3).
\end{split}
\end{align}
This is the three-point analogue of Eq.~\eqref{eq:ILM}: the indices $(L,M)$ carried by $I^{(3)L}_{M}$ are the total angular momentum and azimuthal index of the TripoSH expansion, summed over all internal ranks. As in the two-point case, this is what makes the rotational average economical: $I^{(3)L}_{M}$ is obtained directly as an average over rotated configurations, without having to evaluate the individual TripoSH coefficients $t^{LM}_{\ell_1 \ell_2 \ell_3 \ell_{23}}$, whose internal rank coupling is considerably more involved than in the two-point case.

The behaviour of this rotation average for a statistically isotropic field is straightforward to obtain. Under SI, by definition,  the three-point function is invariant, $\mathcal{T}(\rot \un_1, \rot \un_2, \rot \un_3) = \mathcal{T}(\un_1, \un_2, \un_3)$ for every $\rot \in SO(3)$, so it factors out of the integrand in Eq.~\eqref{eq:I3_LM} and
\begin{align}
\begin{split}
\label{eq:I3_SI}
I^{(3)L}_{M}(\un_1, \un_2, \un_3)
= \mathcal{T}(\un_1, \un_2, \un_3) \int \d\mu(\rot)\, D^{L*}_{MM}(\rot)
= \mathcal{T}(\un_1, \un_2, \un_3)\, \delta_{L0}\, \delta_{M0},
\end{split}
\end{align}
where we have used $D^{0}_{00}(\rot) = 1$ and the orthogonality relation Eq.~\eqref{eq:wigner_ortho}.

This result can also be corroborated in harmonic space, where SI restricts the TripoSH spectrum to a single non-vanishing component, $L=0$, regardless of the internal coupling structure. 
To see this, let us note that statistical isotropy fixes the harmonic three-point function to the form~(\cite{Varshalovich:1988ifq})
\begin{align}
\label{eq:3pt_si}
\Braket{a_{\ell_1 m_1} a_{\ell_2 m_2} a_{\ell_3 m_3}}
= B_{\ell_1 \ell_2 \ell_3}
\begin{pmatrix}
\ell_1 & \ell_2 & \ell_3\\
m_1 & m_2 & m_3
\end{pmatrix},
\end{align}
where $B_{\ell_1 \ell_2 \ell_3}$ is the bispectrum that captures the couplings of the internal ranks. 
In terms of the spherical harmonic coefficients $a_{\ell m}$, the three-point function reads
  \begin{align}   \label{3pcfsh}
    \mathcal{T}(\un_1, \un_2, \un_3)
    = \sum_{\ell_1 \ell_2 \ell_3}\sum_{m_1 m_2 m_3}
    \Braket{a_{\ell_1 m_1} a_{\ell_2 m_2} a_{\ell_3 m_3}}
    Y_{\ell_1 m_1}(\un_1) Y_{\ell_2 m_2}(\un_2) Y_{\ell_3 m_3}(\un_3).
  \end{align}
Projecting Eq.~\eqref{eq:3pcf_triposh} on the TripoSH basis, using Eq~\eqref{3pcfsh} and the orthonormality of the spherical harmonics relates the TripoSH coefficients to the harmonic three-point function,
\begin{align}
\begin{split}
\label{eq:t_from_a}
t_{\ell_1 \ell_2 \ell_3 \ell_{23}}^{LM}
=& \sum_{m_1 m_2 m_3 m_{23}}
\Braket{a_{\ell_1 m_1} a_{\ell_2 m_2} a_{\ell_3 m_3}}
\cg_{\ell_1 m_1 \ell_{23} m_{23}}^{LM}\,
\cg_{\ell_2 m_2 \ell_3 m_3}^{\ell_{23} m_{23}},
\end{split}
\end{align}
which under SI~\eqref{eq:3pt_si} collapses to
\begin{align}
\label{eq:t_si}
t_{\ell_1 \ell_2 \ell_3 \ell_{23}}^{LM}
= (-1)^{\ell_1 + \ell_2 + \ell_3} B_{\ell_1 \ell_2 \ell_3}\,
\delta_{\ell_{23} \ell_1}\, \delta_{L0}\, \delta_{M0}.
\end{align}
The key observation from the above relation is that statistical isotropy populates only the $L = 0$ component of $t_{\ell_1 \ell_2 \ell_3 \ell_{23}}^{LM}$, and thus $I^{(3)L}_{M}$ (through eq.~\ref{eq:I3_t}), regardless of the strength or shape of the non-Gaussianity $B_{\ell_1 \ell_2 \ell_3}$. 

Note that for a zero-mean Gaussian field the three-point function vanishes identically, and with
it every $I^{(3)L}_{M}$, including the $L = 0$ mode. For a non-Gaussian field with a
non-vanishing three-point function, Eq.~\eqref{eq:I3_SI} confines the signal entirely to
$L = 0$ under SI, so that a signal at $L \neq 0$ implies a violation of statistical isotropy in
the three-point function. This is not contained in the two-point measure: for a non-Gaussian
field, statistical isotropy of the two-point function does not imply statistical isotropy of the
field, and $I^{(3)L}_{M}$ can be non-zero at $L \neq 0$ while $I^{L}_{M}$ vanishes there.

The above exercise emphasizes that the real space approach provides a more intuitive and direct way to capture nSI features in the field compared to the harmonic space route, as it does not require computation of the TripoSH coefficients up to arbitrarily high internal ranks.

\bibliographystyle{aasjournalv7}
\bibliography{ref}

@book{Varshalovich:1988ifq,
    author = "Varshalovich, D. A. and Moskalev, A. N. and Khersonskii, V. K.",
    title = "{Quantum Theory of Angular Momentum}: {Irreducible Tensors, Spherical Harmonics, Vector Coupling Coefficients, 3nj Symbols}",
    doi = "10.1142/0270",
    isbn = "978-981-4415-49-1, 978-9971-5-0107-5",
    publisher = "World Scientific Publishing Company",
    year = "1988"
}

@article{Schwarz_2016,
 author = {Schwarz, Dominik J and Copi, Craig J and Huterer, Dragan and Starkman, Glenn D},
 title = {CMB anomalies after Planck},
 journal = {Classical and Quantum Gravity},
 publisher = {IOP Publishing},
 year = {2016},
 month = {aug},
 number = {18},
 volume = {33},
 pages = {184001},
 url = {https://doi.org/10.1088%2F0264-9381%2F33%2F18%2F184001},
 doi = {10.1088/0264-9381/33/18/184001}
}

@article{planck18_VII,
 author = {Akrami, Y. and Ashdown, M. and Aumont, J. and Baccigalupi, C. and Ballardini, M. and Banday, A. J. and Barreiro, R. B. and Bartolo, N. and Basak, S. and Benabed, K. and Bersanelli, M. and Bielewicz, P. and Bock, J. J. and Bond, J. R. and Borrill, J. and Bouchet, F. R. and Boulanger, F. and Bucher, M. and Burigana, C. and Butler, R. C. and Calabrese, E. and Cardoso, J.-F. and Casaponsa, B. and Chiang, H. C. and Colombo, L. P. L. and Combet, C. and Contreras, D. and Crill, B. P. and de Bernardis, P. and de Zotti, G. and Delabrouille, J. and Delouis, J.-M. and Di Valentino, E. and Diego, J. M. and Doré, O. and Douspis, M. and Ducout, A. and Dupac, X. and Efstathiou, G. and Elsner, F. and Enßlin, T. A. and Eriksen, H. K. and Fantaye, Y. and Fernandez-Cobos, R. and Finelli, F. and Frailis, M. and Fraisse, A. A. and Franceschi, E. and Frolov, A. and Galeotta, S. and Galli, S. and Ganga, K. and Génova-Santos, R. T. and Gerbino, M. and Ghosh, T. and González-Nuevo, J. and Górski, K. M. and Gruppuso, A. and Gudmundsson, J. E. and Hamann, J. and Handley, W. and Hansen, F. K. and Herranz, D. and Hivon, E. and Huang, Z. and Jaffe, A. H. and Jones, W. C. and Keihänen, E. and Keskitalo, R. and Kiiveri, K. and Kim, J. and Krachmalnicoff, N. and Kunz, M. and Kurki-Suonio, H. and Lagache, G. and Lamarre, J.-M. and Lasenby, A. and Lattanzi, M. and Lawrence, C. R. and Le Jeune, M. and Levrier, F. and Liguori, M. and Lilje, P. B. and Lindholm, V. and López-Caniego, M. and Ma, Y.-Z. and Macías-Pérez, J. F. and Maggio, G. and Maino, D. and Mandolesi, N. and Mangilli, A. and Marcos-Caballero, A. and Maris, M. and Martin, P. G. and Martínez-González, E. and Matarrese, S. and Mauri, N. and McEwen, J. D. and Meinhold, P. R. and Mennella, A. and Migliaccio, M. and Miville-Deschênes, M.-A. and Molinari, D. and Moneti, A. and Montier, L. and Morgante, G. and Moss, A. and Natoli, P. and Pagano, L. and Paoletti, D. and Partridge, B. and Perrotta, F. and Pettorino, V. and Piacentini, F. and Polenta, G. and Puget, J.-L. and Rachen, J. P. and Reinecke, M. and Remazeilles, M. and Renzi, A. and Rocha, G. and Rosset, C. and Roudier, G. and Rubiño-Martín, J. A. and Ruiz-Granados, B. and Salvati, L. and Savelainen, M. and Scott, D. and Shellard, E. P. S. and Sirignano, C. and Sunyaev, R. and Suur-Uski, A.-S. and Tauber, J. A. and Tavagnacco, D. and Tenti, M. and Toffolatti, L. and Tomasi, M. and Trombetti, T. and Valenziano, L. and Valiviita, J. and Van Tent, B. and Vielva, P. and Villa, F. and Vittorio, N. and Wandelt, B. D. and Wehus, I. K. and Zacchei, A. and Zibin, J. P. and Zonca, A.},
 title = {Planck2018 results: VII. Isotropy and statistics of the CMB},
 journal = {Astronomy \& Astrophysics},
 publisher = {EDP Sciences},
 year = {2020},
 month = {September},
 volume = {641},
 pages = {A7},
 url = {http://dx.doi.org/10.1051/0004-6361/201935201},
 doi = {10.1051/0004-6361/201935201},
 issn = {1432-0746}
}

@article{Shaikh_2019,
   title={Joint Bayesian analysis of large angular scale CMB temperature anomalies},
   volume={2019},
   ISSN={1475-7516},
   url={http://dx.doi.org/10.1088/1475-7516/2019/08/007},
   DOI={10.1088/1475-7516/2019/08/007},
   number={08},
   journal={Journal of Cosmology and Astroparticle Physics},
   publisher={IOP Publishing},
   author={Shaikh, Shabbir and Mukherjee, Suvodip and Das, Santanu and Wandelt, Benjamin D. and Souradeep, Tarun},
   year={2019},
   month=Aug, pages={007–007} }

@article{Planck:2018nkj,
    author = "Aghanim, N. and others",
    collaboration = "Planck",
    title = "{Planck 2018 results. I. Overview and the cosmological legacy of Planck}",
    eprint = "1807.06205",
    archivePrefix = "arXiv",
    primaryClass = "astro-ph.CO",
    doi = "10.1051/0004-6361/201833880",
    journal = "Astron. Astrophys.",
    volume = "641",
    pages = "A1",
    year = "2020"
}

@article{Planck:2015igc,
    author = "Ade, P. A. R. and others",
    collaboration = "Planck",
    title = "{Planck 2015 results. XVI. Isotropy and statistics of the CMB}",
    eprint = "1506.07135",
    archivePrefix = "arXiv",
    primaryClass = "astro-ph.CO",
    doi = "10.1051/0004-6361/201526681",
    journal = "Astron. Astrophys.",
    volume = "594",
    pages = "A16",
    year = "2016"
}

@article{Kumar:2014nda,
    author = "Kumar, Saurabh and Rotti, Aditya and Aich, Moumita and Pant, Nidhi and Mitra, Sanjit and Souradeep, Tarun",
    title = "{Orthogonal bipolar spherical harmonics measures: Scrutinizing sources of isotropy violation}",
    eprint = "1409.4886",
    archivePrefix = "arXiv",
    primaryClass = "astro-ph.CO",
    doi = "10.1103/PhysRevD.91.043501",
    journal = "Phys. Rev. D",
    volume = "91",
    pages = "043501",
    year = "2015"
}

@article{Mukherjee:2013kga,
    author = "Mukherjee, Suvodip and Souradeep, Tarun",
    title = "{Statistically anisotropic Gaussian simulations of the CMB temperature field}",
    eprint = "1311.5837",
    archivePrefix = "arXiv",
    primaryClass = "astro-ph.CO",
    doi = "10.1103/PhysRevD.89.063013",
    journal = "Phys. Rev. D",
    volume = "89",
    number = "6",
    pages = "063013",
    year = "2014"
}

@article{Joshi:2009mj,
    author = "Joshi, Nidhi and Jhingan, S. and Souradeep, Tarun and Hajian, Amir",
    title = "{Bipolar Harmonic encoding of CMB correlation patterns}",
    eprint = "0912.3217",
    archivePrefix = "arXiv",
    primaryClass = "astro-ph.CO",
    doi = "10.1103/PhysRevD.81.083012",
    journal = "Phys. Rev. D",
    volume = "81",
    pages = "083012",
    year = "2010"
}

@article{Mitra:2004nx,
    author = "Mitra, Sanjit and Sengupta, Anand S. and Souradeep, Tarun",
    title = "{CMB power spectrum estimation using non-circular beam}",
    eprint = "astro-ph/0405406",
    archivePrefix = "arXiv",
    reportNumber = "IUCAA-14-2004",
    doi = "10.1103/PhysRevD.70.103002",
    journal = "Phys. Rev. D",
    volume = "70",
    pages = "103002",
    year = "2004"
}

@article{Hajian:2003qq,
    author = "Hajian, Amir and Souradeep, Tarun",
    title = "{Measuring statistical isotropy of the CMB anisotropy}",
    eprint = "astro-ph/0308001",
    archivePrefix = "arXiv",
    reportNumber = "IUCAA-37-2003",
    doi = "10.1086/379757",
    journal = "Astrophys. J. Lett.",
    volume = "597",
    pages = "L5--L8",
    year = "2003"
}

@article{Dipanshu:2023tix,
    author = "Dipanshu and Souradeep, Tarun and Hirve, Shriya",
    title = "{Capturing Statistical Isotropy Violation with Generalized Isotropic Angular Correlation Functions of Cosmic Microwave Background Anisotropy}",
    eprint = "2301.04539",
    archivePrefix = "arXiv",
    primaryClass = "astro-ph.CO",
    doi = "10.3847/1538-4357/ace895",
    journal = "Astrophys. J.",
    volume = "954",
    number = "2",
    pages = "181",
    year = "2023"
}

@article{Das:2014awa,
    author = "Das, Santanu and Mitra, Sanjit and Rotti, Aditya and Pant, Nidhi and Souradeep, Tarun",
    title = "{Statistical isotropy violation in WMAP CMB maps resulting from non-circular beams}",
    eprint = "1401.7757",
    archivePrefix = "arXiv",
    primaryClass = "astro-ph.CO",
    reportNumber = "IUCAA-03-2014",
    doi = "10.1051/0004-6361/201424164",
    journal = "Astron. Astrophys.",
    volume = "591",
    pages = "A97",
    year = "2016"
}

@article{Prunet:2004zy,
    author = "Prunet, Simon and Uzan, Jean-Philippe and Bernardeau, Francis and Brunier, Tristan",
    title = "{Constraints on mode couplings and modulation of the CMB with WMAP data}",
    eprint = "astro-ph/0406364",
    archivePrefix = "arXiv",
    doi = "10.1103/PhysRevD.71.083508",
    journal = "Phys. Rev. D",
    volume = "71",
    pages = "083508",
    year = "2005"
}

@article{Gordon_2007,
 author = {Gordon, Christopher},
 title = {Broken Isotropy from a Linear Modulation of the Primordial Perturbations},
 journal = {The Astrophysical Journal},
 publisher = {American Astronomical Society},
 year = {2007},
 month = {February},
 number = {2},
 volume = {656},
 pages = {636--640},
 url = {http://dx.doi.org/10.1086/510511},
 doi = {10.1086/510511},
 issn = {1538-4357}
}

@article{Carroll:2008br,
    author = "Carroll, Sean M. and Tseng, Chien-Yao and Wise, Mark B.",
    title = "{Translational Invariance and the Anisotropy of the Cosmic Microwave Background}",
    eprint = "0811.1086",
    archivePrefix = "arXiv",
    primaryClass = "astro-ph",
    reportNumber = "CALT-68-2705",
    doi = "10.1103/PhysRevD.81.083501",
    journal = "Phys. Rev. D",
    volume = "81",
    pages = "083501",
    year = "2010"
}

@article{Bond_1998,
   title={Computing CMB anisotropy in compact hyperbolic spaces},
   volume={15},
   ISSN={1361-6382},
   url={http://dx.doi.org/10.1088/0264-9381/15/9/014},
   DOI={10.1088/0264-9381/15/9/014},
   number={9},
   journal={Classical and Quantum Gravity},
   publisher={IOP Publishing},
   author={Bond, J Richard and Pogosyan, Dmitry and Souradeep, Tarun},
   year={1998},
   month=Sept, pages={2671–2687} }

@article{Planck_2013,
   title={<i>Planck</i>2013 results. XXIII. Isotropy and statistics of the CMB},
   volume={571},
   ISSN={1432-0746},
   url={http://dx.doi.org/10.1051/0004-6361/201321534},
   DOI={10.1051/0004-6361/201321534},
   journal={Astronomy \&; Astrophysics},
   publisher={EDP Sciences},
   author={{Planck Collaboration} and Ade, P. A. R. and Aghanim, N. and Armitage-Caplan, C. and Arnaud, M. and Ashdown, M. and Atrio-Barandela, F. and Aumont, J. and Baccigalupi, C. and Banday, A. J. and Barreiro, R. B. and Bartlett, J. G. and Bartolo, N. and Battaner, E. and Battye, R. and Benabed, K. and Benoît, A. and Benoit-Lévy, A. and Bernard, J.-P. and Bersanelli, M. and Bielewicz, P. and Bobin, J. and Bock, J. J. and Bonaldi, A. and Bonavera, L. and Bond, J. R. and Borrill, J. and Bouchet, F. R. and Bridges, M. and Bucher, M. and Burigana, C. and Butler, R. C. and Cardoso, J.-F. and Catalano, A. and Challinor, A. and Chamballu, A. and Chary, R.-R. and Chiang, H. C. and Chiang, L.-Y and Christensen, P. R. and Church, S. and Clements, D. L. and Colombi, S. and Colombo, L. P. L. and Couchot, F. and Coulais, A. and Crill, B. P. and Cruz, M. and Curto, A. and Cuttaia, F. and Danese, L. and Davies, R. D. and Davis, R. J. and de Bernardis, P. and de Rosa, A. and de Zotti, G. and Delabrouille, J. and Delouis, J.-M. and Désert, F.-X. and Diego, J. M. and Dole, H. and Donzelli, S. and Doré, O. and Douspis, M. and Ducout, A. and Dupac, X. and Efstathiou, G. and Elsner, F. and Enßlin, T. A. and Eriksen, H. K. and Fantaye, Y. and Fergusson, J. and Finelli, F. and Forni, O. and Frailis, M. and Franceschi, E. and Frommert, M. and Galeotta, S. and Ganga, K. and Giard, M. and Giardino, G. and Giraud-Héraud, Y. and González-Nuevo, J. and Górski, K. M. and Gratton, S. and Gregorio, A. and Gruppuso, A. and Hansen, F. K. and Hansen, M. and Hanson, D. and Harrison, D. L. and Helou, G. and Henrot-Versillé, S. and Hernández-Monteagudo, C. and Herranz, D. and Hildebrandt, S. R. and Hivon, E. and Hobson, M. and Holmes, W. A. and Hornstrup, A. and Hovest, W. and Huffenberger, K. M. and Jaffe, A. H. and Jaffe, T. R. and Jones, W. C. and Juvela, M. and Keihänen, E. and Keskitalo, R. and Kim, J. and Kisner, T. S. and Knoche, J. and Knox, L. and Kunz, M. and Kurki-Suonio, H. and Lagache, G. and Lähteenmäki, A. and Lamarre, J.-M. and Lasenby, A. and Laureijs, R. J. and Lawrence, C. R. and Leahy, J. P. and Leonardi, R. and Leroy, C. and Lesgourgues, J. and Liguori, M. and Lilje, P. B. and Linden-Vørnle, M. and López-Caniego, M. and Lubin, P. M. and Macías-Pérez, J. F. and Maffei, B. and Maino, D. and Mandolesi, N. and Mangilli, A. and Marinucci, D. and Maris, M. and Marshall, D. J. and Martin, P. G. and Martínez-González, E. and Masi, S. and Massardi, M. and Matarrese, S. and Matthai, F. and Mazzotta, P. and McEwen, J. D. and Meinhold, P. R. and Melchiorri, A. and Mendes, L. and Mennella, A. and Migliaccio, M. and Mikkelsen, K. and Mitra, S. and Miville-Deschênes, M.-A. and Molinari, D. and Moneti, A. and Montier, L. and Morgante, G. and Mortlock, D. and Moss, A. and Munshi, D. and Murphy, J. A. and Naselsky, P. and Nati, F. and Natoli, P. and Netterfield, C. B. and Nørgaard-Nielsen, H. U. and Noviello, F. and Novikov, D. and Novikov, I. and Osborne, S. and Oxborrow, C. A. and Paci, F. and Pagano, L. and Pajot, F. and Paoletti, D. and Pasian, F. and Patanchon, G. and Peiris, H. V. and Perdereau, O. and Perotto, L. and Perrotta, F. and Piacentini, F. and Piat, M. and Pierpaoli, E. and Pietrobon, D. and Plaszczynski, S. and Pogosyan, D. and Pointecouteau, E. and Polenta, G. and Ponthieu, N. and Popa, L. and Poutanen, T. and Pratt, G. W. and Prézeau, G. and Prunet, S. and Puget, J.-L. and Rachen, J. P. and Racine, B. and Räth, C. and Rebolo, R. and Reinecke, M. and Remazeilles, M. and Renault, C. and Renzi, A. and Ricciardi, S. and Riller, T. and Ristorcelli, I. and Rocha, G. and Rosset, C. and Rotti, A. and Roudier, G. and Rubiño-Martín, J. A. and Ruiz-Granados, B. and Rusholme, B. and Sandri, M. and Santos, D. and Savini, G. and Scott, D. and Seiffert, M. D. and Shellard, E. P. S. and Souradeep, T. and Spencer, L. D. and Starck, J.-L. and Stolyarov, V. and Stompor, R. and Sudiwala, R. and Sureau, F. and Sutter, P. and Sutton, D. and Suur-Uski, A.-S. and Sygnet, J.-F. and Tauber, J. A. and Tavagnacco, D. and Terenzi, L. and Toffolatti, L. and Tomasi, M. and Tristram, M. and Tucci, M. and Tuovinen, J. and Türler, M. and Valenziano, L. and Valiviita, J. and Van Tent, B. and Varis, J. and Vielva, P. and Villa, F. and Vittorio, N. and Wade, L. A. and Wandelt, B. D. and Wehus, I. K. and White, M. and Wilkinson, A. and Yvon, D. and Zacchei, A. and Zonca, A.},
   year={2014},
   month=Oct, pages={A23} }

@article{Planck_2018,
   title={<i>Planck</i>2018 results: VII. Isotropy and statistics of the CMB},
   volume={641},
   ISSN={1432-0746},
   url={http://dx.doi.org/10.1051/0004-6361/201935201},
   DOI={10.1051/0004-6361/201935201},
   journal={Astronomy \&; Astrophysics},
   publisher={EDP Sciences},
   author={{Planck Collaboration} and Akrami, Y. and Ashdown, M. and Aumont, J. and Baccigalupi, C. and Ballardini, M. and Banday, A. J. and Barreiro, R. B. and Bartolo, N. and Basak, S. and Benabed, K. and Bersanelli, M. and Bielewicz, P. and Bock, J. J. and Bond, J. R. and Borrill, J. and Bouchet, F. R. and Boulanger, F. and Bucher, M. and Burigana, C. and Butler, R. C. and Calabrese, E. and Cardoso, J.-F. and Casaponsa, B. and Chiang, H. C. and Colombo, L. P. L. and Combet, C. and Contreras, D. and Crill, B. P. and de Bernardis, P. and de Zotti, G. and Delabrouille, J. and Delouis, J.-M. and Di Valentino, E. and Diego, J. M. and Doré, O. and Douspis, M. and Ducout, A. and Dupac, X. and Efstathiou, G. and Elsner, F. and Enßlin, T. A. and Eriksen, H. K. and Fantaye, Y. and Fernandez-Cobos, R. and Finelli, F. and Frailis, M. and Fraisse, A. A. and Franceschi, E. and Frolov, A. and Galeotta, S. and Galli, S. and Ganga, K. and Génova-Santos, R. T. and Gerbino, M. and Ghosh, T. and González-Nuevo, J. and Górski, K. M. and Gruppuso, A. and Gudmundsson, J. E. and Hamann, J. and Handley, W. and Hansen, F. K. and Herranz, D. and Hivon, E. and Huang, Z. and Jaffe, A. H. and Jones, W. C. and Keihänen, E. and Keskitalo, R. and Kiiveri, K. and Kim, J. and Krachmalnicoff, N. and Kunz, M. and Kurki-Suonio, H. and Lagache, G. and Lamarre, J.-M. and Lasenby, A. and Lattanzi, M. and Lawrence, C. R. and Le Jeune, M. and Levrier, F. and Liguori, M. and Lilje, P. B. and Lindholm, V. and López-Caniego, M. and Ma, Y.-Z. and Macías-Pérez, J. F. and Maggio, G. and Maino, D. and Mandolesi, N. and Mangilli, A. and Marcos-Caballero, A. and Maris, M. and Martin, P. G. and Martínez-González, E. and Matarrese, S. and Mauri, N. and McEwen, J. D. and Meinhold, P. R. and Mennella, A. and Migliaccio, M. and Miville-Deschênes, M.-A. and Molinari, D. and Moneti, A. and Montier, L. and Morgante, G. and Moss, A. and Natoli, P. and Pagano, L. and Paoletti, D. and Partridge, B. and Perrotta, F. and Pettorino, V. and Piacentini, F. and Polenta, G. and Puget, J.-L. and Rachen, J. P. and Reinecke, M. and Remazeilles, M. and Renzi, A. and Rocha, G. and Rosset, C. and Roudier, G. and Rubiño-Martín, J. A. and Ruiz-Granados, B. and Salvati, L. and Savelainen, M. and Scott, D. and Shellard, E. P. S. and Sirignano, C. and Sunyaev, R. and Suur-Uski, A.-S. and Tauber, J. A. and Tavagnacco, D. and Tenti, M. and Toffolatti, L. and Tomasi, M. and Trombetti, T. and Valenziano, L. and Valiviita, J. and Van Tent, B. and Vielva, P. and Villa, F. and Vittorio, N. and Wandelt, B. D. and Wehus, I. K. and Zacchei, A. and Zibin, J. P. and Zonca, A.},
   year={2020},
   month=Sept, pages={A7} }

@article{Bartolo_2012,
 author = {Bartolo, N and Dimastrogiovanni, E and Liguori, M and Matarrese, S and Riotto, A},
 title = {An estimator for statistical anisotropy from the CMB bispectrum},
 journal = {Journal of Cosmology and Astroparticle Physics},
 publisher = {IOP Publishing},
 year = {2012},
 month = {January},
 number = {01},
 volume = {2012},
 pages = {029--029},
 url = {http://dx.doi.org/10.1088/1475-7516/2012/01/029},
 doi = {10.1088/1475-7516/2012/01/029},
 issn = {1475-7516}
}

@article{Rath_2015,
 author = {Rath, Pranati K. and Aluri, Pavan K. and Jain, Pankaj},
 title = {Relating the inhomogeneous power spectrum to the CMB hemispherical anisotropy},
 journal = {Physical Review D},
 publisher = {American Physical Society (APS)},
 year = {2015},
 month = {jan},
 number = {2},
 volume = {91},
 url = {https://doi.org/10.1103%2Fphysrevd.91.023515},
 doi = {10.1103/physrevd.91.023515}
}

@article{Kothari_2016,
 author = {Kothari, Rahul and Ghosh, Shamik and Rath, Pranati K. and Kashyap, Gopal and Jain, Pankaj},
 title = {Imprint of inhomogeneous and anisotropic primordial power spectrum on CMB polarization},
 journal = {Monthly Notices of the Royal Astronomical Society},
 publisher = {Oxford University Press (OUP)},
 year = {2016},
 month = {may},
 number = {2},
 volume = {460},
 pages = {1577--1587},
 url = {https://doi.org/10.1093%2Fmnras%2Fstw1039},
 doi = {10.1093/mnras/stw1039}
}

@article{Ragavendra_2025,
 author = {Ragavendra, H. V. and Mukherjee, Dipayan and Sethi, Shiv K.},
 title = {Cosmological consequences of statistical inhomogeneity},
 journal = {Physical Review D},
 publisher = {American Physical Society (APS)},
 year = {2025},
 month = {January},
 number = {2},
 volume = {111},
 url = {http://dx.doi.org/10.1103/PhysRevD.111.023541},
 doi = {10.1103/physrevd.111.023541},
 issn = {2470-0029}
}

@misc{Hajian:2005jh,
    author = "Hajian, Amir and Souradeep, Tarun",
    title = "{The Cosmic microwave background bipolar power spectrum: Basic formalism and applications}",
    eprint = "astro-ph/0501001",
    archivePrefix = "arXiv",
    month = "1",
    year = "2005"
}

@article{Hajian:2006ud,
    author = "Hajian, Amir and Souradeep, Tarun",
    title = "{Testing Global Isotropy of Three-Year Wilkinson Microwave Anisotropy Probe (WMAP) Data: Temperature Analysis}",
    eprint = "astro-ph/0607153",
    archivePrefix = "arXiv",
    doi = "10.1103/PhysRevD.74.123521",
    journal = "Phys. Rev. D",
    volume = "74",
    pages = "123521",
    year = "2006"
}

@article{Wu:1998ad,
    author = "Wu, Kelvin K. S. and Lahav, Ofer and Rees, Martin J.",
    editor = "Sato, H. and Sugiyama, N.",
    title = "{The large-scale smoothness of the Universe}",
    eprint = "astro-ph/9804062",
    archivePrefix = "arXiv",
    reportNumber = "OL-07-98",
    doi = "10.1038/16637",
    journal = "Nature",
    volume = "397",
    pages = "225--230",
    year = "1999"
}

@article{Hivon:2001jp,
    author = "Hivon, E. and Gorski, K. M. and Netterfield, C. B. and Crill, B. P. and Prunet, S. and Hansen, F.",
    title = "{Master of the cosmic microwave background anisotropy power spectrum: a fast method for statistical analysis of large and complex cosmic microwave background data sets}",
    eprint = "astro-ph/0105302",
    archivePrefix = "arXiv",
    doi = "10.1086/338126",
    journal = "Astrophys. J.",
    volume = "567",
    pages = "2",
    year = "2002"
}

@article{Lewis:2001hp,
    author = "Lewis, Antony and Challinor, Anthony and Turok, Neil",
    title = "{Analysis of CMB polarization on an incomplete sky}",
    eprint = "astro-ph/0106536",
    archivePrefix = "arXiv",
    doi = "10.1103/PhysRevD.65.023505",
    journal = "Phys. Rev. D",
    volume = "65",
    pages = "023505",
    year = "2002"
}

@article{Carvalho:2010rz,
    author = "Carvalho, C. S. and Moodley, K.",
    title = "{Real space estimator for the weak lensing convergence from the CMB}",
    eprint = "1005.4288",
    archivePrefix = "arXiv",
    primaryClass = "astro-ph.CO",
    doi = "10.1103/PhysRevD.81.123010",
    journal = "Phys. Rev. D",
    volume = "81",
    pages = "123010",
    year = "2010"
}

@article{Bucher:2010iv,
    author = "Bucher, Martin and Carvalho, Carla Sofia and Moodley, Kavilan and Remazeilles, Mathieu",
    title = "{CMB Lensing Reconstruction in Real Space}",
    eprint = "1004.3285",
    archivePrefix = "arXiv",
    primaryClass = "astro-ph.CO",
    doi = "10.1103/PhysRevD.85.043016",
    journal = "Phys. Rev. D",
    volume = "85",
    pages = "043016",
    year = "2012"
}

@misc{Szapudi:2001qj,
    author = "Szapudi, I. and Prunet, S. and Colombi, S.",
    title = "{Fast clustering analysis of inhomogeneous megapixel cmb maps}",
    eprint = "astro-ph/0107383",
    archivePrefix = "arXiv",
    month = "7",
    year = "2001"
}

@book{shoemake,
    author = "Shoemake, Ken",
    title = "{Graphic Gems III}",
    editor = "Kirk, David",
    publisher = "Academic Press",
    pages = {124--132},
    year = "1992",
    doi  = {10.1016/B978-0-08-050755-2.50036-1}
}

@article{Friar:1981zza,
    author = "Friar, James Lewis and Tomusiak, E. L. and Gibson, Benjamin F. and Payne, G. L.",
    title = "{Faddeev wave function decomposition using bipolar harmonics}",
    doi = "10.1103/PhysRevC.24.677",
    journal = "Phys. Rev. C",
    volume = "24",
    pages = "677--683",
    year = "1981"
}

@article{Ershov:2004va,
    author = "Ershov, S. N.",
    title = "{General structure of a two-body operator for spin- 1 2 particles}",
    doi = "10.1103/PhysRevC.70.054604",
    journal = "Phys. Rev. C",
    volume = "70",
    pages = "054604",
    year = "2004"
}

@article{manakov_multipole_2002,
	title = {Multipole expansions of irreducible tensor sets and some applications},
	volume = {35},
	issn = {0953-4075, 1361-6455},
	url = {https://iopscience.iop.org/article/10.1088/0953-4075/35/1/306},
	doi = {10.1088/0953-4075/35/1/306},
	number = {1},
	urldate = {2026-05-05},
	journal = {Journal of Physics B: Atomic, Molecular and Optical Physics},
	author = {Manakov, N L and Meremianin, A V and Starace, Anthony F},
	month = jan,
	year = {2002},
	pages = {77--91},
}

@article{meremianin_three-body_2005,
	title = {The three-body rigid rotator and multipole expansions of the three-body continuum},
	volume = {38},
	issn = {0953-4075, 1361-6455},
	url = {https://iopscience.iop.org/article/10.1088/0953-4075/38/6/013},
	doi = {10.1088/0953-4075/38/6/013},
	number = {6},
	urldate = {2026-05-05},
	journal = {Journal of Physics B: Atomic, Molecular and Optical Physics},
	author = {Meremianin, A V},
	month = mar,
	year = {2005},
	pages = {757--775},
}

@article{Lachieze-Rey:1995qrb,
    author = "Lachieze-Rey, Marc and Luminet, Jean-Pierre",
    title = "{Cosmic topology}",
    eprint = "gr-qc/9605010",
    archivePrefix = "arXiv",
    doi = "10.1016/0370-1573(94)00085-H",
    journal = "Phys. Rept.",
    volume = "254",
    pages = "135--214",
    year = "1995"
}

@article{Aluri:2022hzs,
    author = "Aluri, Pavan Kumar and others",
    title = "{Is the observable Universe consistent with the cosmological principle?}",
    eprint = "2207.05765",
    archivePrefix = "arXiv",
    primaryClass = "astro-ph.CO",
    doi = "10.1088/1361-6382/acbefc",
    journal = "Class. Quant. Grav.",
    volume = "40",
    number = "9",
    pages = "094001",
    year = "2023"
}

@article{Gorski_2005,
doi = {10.1086/427976},
url = {https://doi.org/10.1086/427976},
year = {2005},
month = {apr},
volume = {622},
number = {2},
pages = {759},
author = {Górski, K. M. and Hivon, E. and Banday, A. J. and Wandelt, B. D. and Hansen, F. K. and Reinecke, M. and Bartelmann, M.},
title = {HEALPix: A Framework for High-Resolution Discretization and Fast Analysis of Data Distributed on the Sphere},
journal = {The Astrophysical Journal}
}

\end{document}